\documentclass[a4paper]{jpconf}
\usepackage{graphicx}
\usepackage{cancel}
\begin{document}
\title{Jet Measurements in Heavy-ion Collisions with CMS}

\author{Raghav Kunnawalkam Elayavalli on behalf of the CMS collaboration}

\address{Rutgers University, Piscataway, USA 08854}

\ead{raghav.k.e@cern.ch}

\begin{abstract}
	The production of a strongly interacting medium in heavy-ion collisions is identified through suppression of high transverse momentum jets leading to an effect known as jet quenching. Detailed measurements of nuclear modification factors and energy flow of the aforementioned quenched jets are performed with the CMS detector. These new results extend previous measurements to large angles with respect to the quenched jets and increase reach in the transverse momentum range. High statistics pp, pPb and PbPb data taken in 2011-13 are used in the following analysis. A new data-driven method to estimate the underlying event contribution from the forward calorimeter energy distribution is employed which takes into account possible event by event flow modulation. The flavor dependence of energy loss of jets is studied by tagging heavy flavor b-jets in these systems.
	\\
	\\
	Presented at {\it Hot Quarks 2014}
\end{abstract}

\section{Introduction}

	Relativistic heavy-ions with sufficient energy density are expected to produce a strongly interacting medium known as the quark gluon plasma (QGP)~\cite{QGP}. Large transverse momentum ($p_T$) jets are of a special importance because their medium interactions lead to a modification in their yield, i.e. they are quenched~\cite{jet_quenching} when compared to pp collisions at the same energy. This was seen at RHIC in central AuAu collisions~\cite{rhic_quenching1,rhic_quenching2,rhic_quenching3,rhic_quenching4} and at the LHC~\cite{12004}. Recent datasets with integrated luminosity $150 ~\mu b^{-1}$ in PbPb collisions at $\sqrt{s_{NN}}=2.76$ TeV and $35 ~nb^{-1}$ in pPb collisions at $\sqrt{s_{NN}}=5.02$ TeV. A reference pp dataset with integrated luminosity ($5.3~pb^{-1}$ at $\sqrt{s}=2.76$ TeV) is used for the calculation of the nuclear modification factor. 
	
	The jet quenching effect is assumed to be dependent on the flavor of the fragmenting quark. For example when the parton $p_T$ is comparable with the parton mass, the radiation is expected to be less due to the dead-cone effect~\cite{dead_cone}. Heavy flavor quark jets are tagged using a discriminant based on the displaced vertex and the b-jet yields are compared between the three systems to study the flavor dependence of quenching. A novel method of studying the distribution of track $p_T$ around the jet axis for a particular jet (in a dijet system) is performed to recover the energy lost due to quenching by going to large jet radii in the $\phi, \eta$ plane. The detector coordinate system is defined with $\phi$ as the azimuthal angle and $\eta$, the pseudorapidity. The differences in charged particle multiplicity and momentum flow carried by charged particles near the leading and the sub-leading jets in dijet events are measured in PbPb and pp systems. 

	Events in heavy-ion collisions are also classified according to their centrality or the degree of overlap of the two colliding nuclei. Events with a very large impact parameter correspond to a small overlap region and hence are termed as peripheral events. A small impact parameter pictorially represents two Lorentz contracted nuclei with a head-on collision and these are termed as most central collisions. The centrality classes are separated based on the energy deposited in the Hadronic Forward (HF) calorimeters for the PbPb events and by looking at the number of reconstructed charged particle tracks (event multiplicity) in the pPb system. 

\section{Jet reconstruction}
	
	A full description of the CMS experiment is available elsewhere~\cite{CMS}. Jets are reconstructed using the CMS particle flow algorithm~\cite{particleflow} which combines all stable particle signatures like calorimeter deposits, track information and muon chamber hits. These particle flow objects are then clustered using the anti-$k_T$ algorithm~\cite{antikt} encoded in the FastJet~\cite{fastjet2} framework. The clustering algorithms use a resolution parameter $R$, which is a $p_T$ weighted distance between objects in the $\eta$ and $\phi$ space to specify the ``size" of the reconstructed jet. To study the momentum flow in dijet measurements, calorimetric jets that are reconstructed from energy deposits in the calorimeters are used. 
	A discriminator called the Simple Secondary Vertex tagger (SSV)~\cite{ssv} which is based on the flight distance significance of the reconstructed secondary vertices is used to tag heavy flavor jets. The Underlying Event (UE) contribution or the background energy is removed using an iterative-pileup subtraction technique~\cite{pu_subtraction}. The HF/Voronoi algorithm~\cite{hf_voronoi} is a new background subtraction method which models the central underlying event density as a nonlinear response of the HF azimuthal profile. This improves the estimation of the central UE by incorporating several non-zero components of the anisotropic flow of the QGP. The estimated energy is subtracted on particle level and is later equalized to even out non physical negative energies (which arise due to the subtraction).  
	
\section{Nuclear modification factors}
	
	One of the important jet tomography variables is the nuclear modification factor ($R_{AA}$ or $R_{pA}$ depending on the collision system) defined as $$ R_{AA} = \frac{dN^{AA}_{jets}/d p_T d\eta}{T_{AA} \cdot d\sigma^{pp}_{jets}/d p_T d\eta}.$$ Here $T_{AA} = N_{coll}/\sigma_{inel}^{NN}$ is the nuclear overlap function defined in terms of the inelastic nucleon-nucleon cross section ($\sigma_{inel}^{NN}$) and the effective number of binary nucleon nucleon collisions ($N_{coll}$) calculated from a Glauber model~\cite{glauber}.   
	
\begin{figure}[h] 
	   \centering
	   \includegraphics[trim=0.25cm 0cm 0cm 0cm, clip=true, scale=0.55]{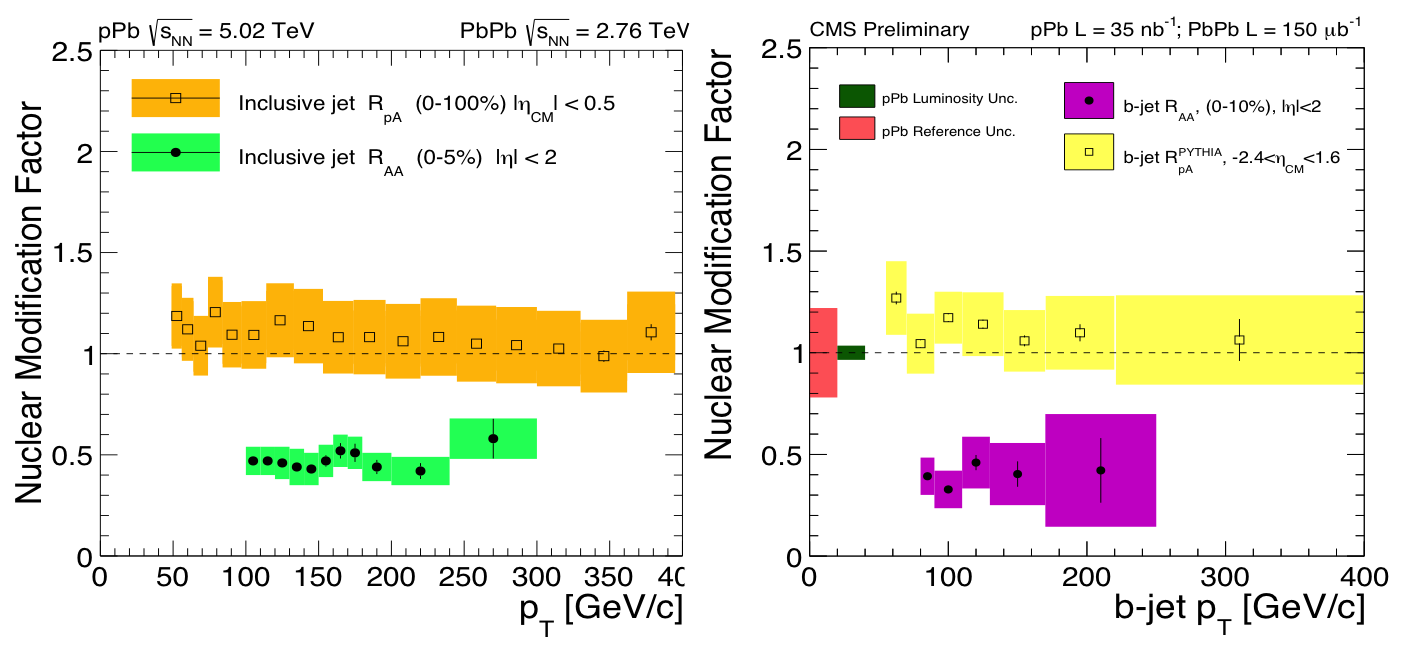} 
	   \caption{Left: Inclusive jet ($R=0.3$, anti-$k_T$ algorithm) nuclear modification factor $R_{AA}$ (closed circles) in the most central PbPb events and $R_{pPb}$ (open squares). Right: b-jet nuclear modification factor measured in pPb (open squares) and central PbPb (closed circles) collisions. Statistical uncertainty is represented by bars and systematics by boxes. Luminosity uncertainty is indicated by the boxes at b-jet $p_T=0$.}
	   \label{fig:raa_rpa}
	\end{figure}

	The nuclear modification factor for fully reconstructed anti-$k_T$ jets in PbPb collisions (closed squares)~\cite{12004} shown in Fig.~\ref{fig:raa_rpa} (left) is compared with the bottom quark tagged jets~\cite{12003} (b-jets) on the right. Jet suppression is found to be flavor independent at high $p_T$ and in central PbPb events. The jets are also suppressed by a factor of about $0.5$ independent of the jet transverse momentum. The $R_{pA}$ is shown in open squares for inclusive jets on the left~\cite{14001} and b-jets on the right~\cite{14007}. Within the uncertainties no deviation from unity is observed. The large uncertainties in the $R_{pA}$ stem from the pp reference extrapolated from published  inclusive jet pp spectra at $\sqrt{s}=7$ TeV using the anti-$k_T$ algorithm with the resolution parameter of $R=0.5,0.7$~\cite{pp_r5,pp_r7}. The center of mass energy dependence is calculated by the ratio of the simulated spectra at $\sqrt{s}=5.02$ to $\sqrt{s}=7$ TeV and was used to scale down the measured pp spectra. A detailed description of the jet measurement and the reference extrapolation can be found in~\cite{14001}.  

	\break

\section{Regaining lost energy}Ê    

\begin{figure}[h!] 
	   \centering
	   \includegraphics[scale=0.25]{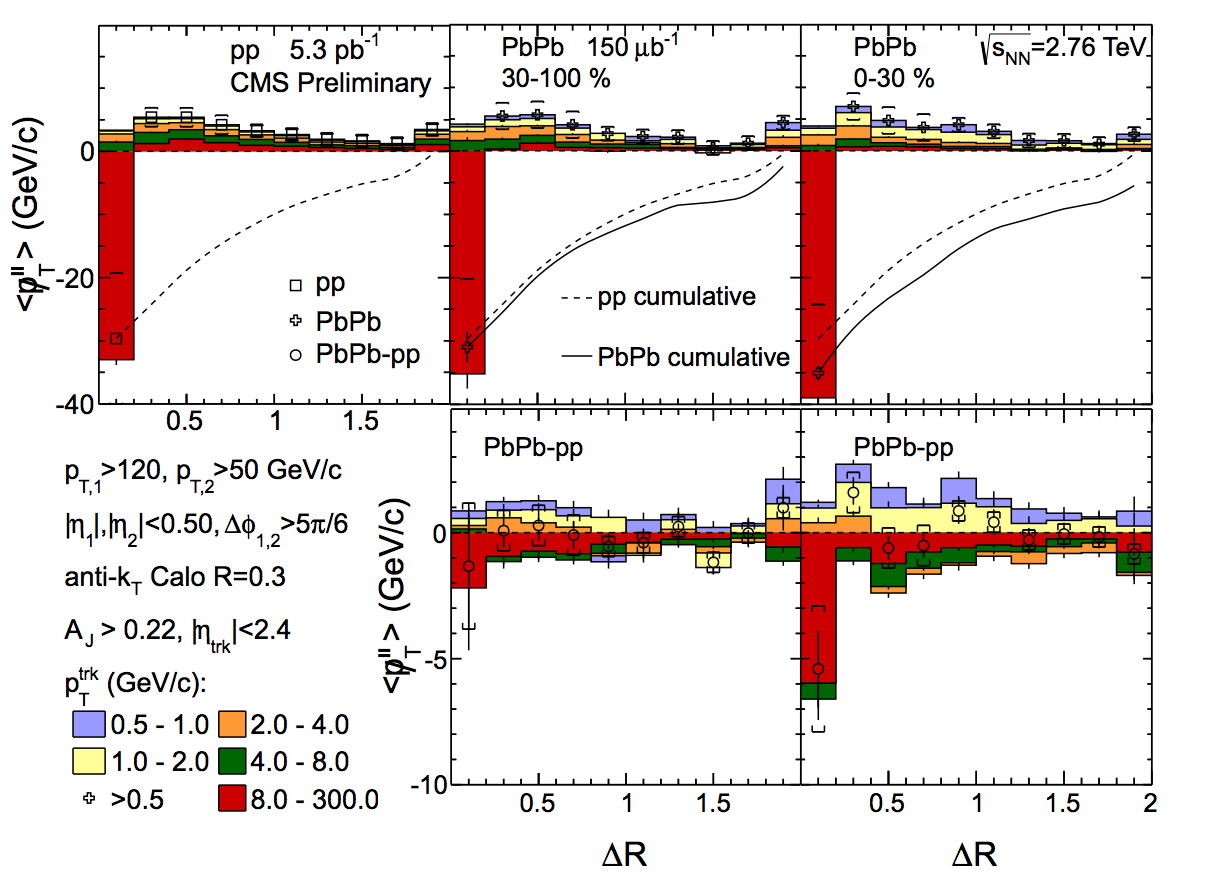} 
	   \caption{Upper row: Differential missing $p_T$ distributions for pp, 30-100\% and 0-30\% PbPb data for various $p_T$ ranges (colored boxes), as a function of $\Delta R$. Also shown is the total $<\cancel{p}_T ^{||}>$ for pp (open boxes), PbPb data (cross). Dashed lines indicate the cumulative $<\cancel{p}_T ^{||}>$ for events with large dijet asymmetry. Bottom row: Difference between PbPb and pp where the bars and brackets represent statistical and systematic uncertainties respectively.}
	   \label{fig:missingpt}
	\end{figure}

	To study dijets where one of the jets is quenched, the dijet asymmetry ratio $A_J = (p_{T,1} - p_{T,2})/(p_{T,1} + p_{T,2})$ is calculated and the most unbalanced dijets (with large $A_J > 0.22$) are selected~\cite{14010}. The direction of the dijet is defined as  $\phi_{Dijet} = \frac{1}{2}(\phi_1 + (\pi - \phi_2))$. The missing $p_T$, defined as $$\cancel{p}_T ^{||} = \sum_{i} -p_T ^i \cos{(\phi_i - \phi_{Dijet})}$$ is used to study the angular flow in rings of $\Delta R = \sqrt{\Delta \phi^2_{trk,Jet} + \Delta \eta^2_{trk,Jet}}$. The top panels of Fig.~\ref{fig:missingpt} show the missing $p_T$ measurement in central PbPb  on the right, peripheral PbPb collisions in the middle and pp collisions on the left. The bottom panels show the difference between PbPb and pp for the central and peripheral collisions on the right and center respectively. The missing $p_T$ is shown as stacked colored histograms (color online) differentiated based on the $p_T$ and the lines, dotted and solid represent the cumulative distribution for pp and PbPb collisions. For large $\Delta R \approx 2$, the missing $p_T$ gets balanced by low $p_T$ particles which is found to be similar for PbPb and pp collisions. 

\section{Conclusions}

	Nuclear modification factors measured for inclusive and b-jets in central PbPb collisions show evidence of strong suppression and are jet flavor independent. Extending this measurement to lower $p_T$ for the heavy flavor jets might highlight the dependence (if any exists) since mass becomes an important factor in that kinematic regime. In contrast to the PbPb measurement, the pPb modification factor does not show any enhancement or suppression in the $p_T$ range reconstructed for inclusive jets and b-jets. To further quantify the energy loss in the quenched jets, the momentum flow of tracks in unbalanced dijet pairs in both PbPb and pp collisions is studied. There is about 35 GeV/c of $p_T$ missing from the most quenched jet in these asymmetric back to back dijet pairs. This loss in $p_T$ is recovered by collecting low $p_T \leq 2$ GeV/c tracks up to $\Delta R \approx 2$ around the dijet axis. The angular pattern of the momentum flow is found to be similar between PbPb and pp while the fragmentation is inherently different between pp and central PbPb collisions.

\end{document}